\documentclass[%
reprint,
twocolumn,
groupedaddress,
 amsmath,amssymb,
 aps, physrev,
nofootinbib,
]{revtex4-2}

\usepackage{graphicx}
\usepackage{dcolumn}
\usepackage{bm}
\usepackage{comment}
\usepackage[dvipsnames]{xcolor}
\usepackage{url}
\usepackage{hyperref}
\usepackage{mhchem}
\usepackage[normalem]{ulem}

\begin{document}

\title{Structural Decomposition of UV–Visible Spectral Variation: Azobenzene in Ethanol Solution}

\author{Eemeli A. Eronen}
\email{eemeli.a.eronen@utu.fi}
\affiliation{University of Turku, Department of Physics and Astronomy, FI-20014 Turun yliopisto, Finland}

\author{Johannes Niskanen}
\email{johannes.niskanen@utu.fi}
\affiliation{University of Turku, Department of Physics and Astronomy, FI-20014 Turun yliopisto, Finland}

\date{\today}

\begin{abstract}
We present a structural interpretation of statistical variability in simulated liquid-phase UV–visible absorption spectra. We analyze the significant variation of the spectral response, caused by structural variation within the ensemble, using a response‑targeted method known as emulator‑based component analysis. In the high-dimensional input space, the method identifies a subspace of a few dimensions that accounts for most spectral variance. The resulting decomposition reveals the spectrally decisive structural features and filters out the irrelevant ones. For our test case, the ethanolic {\it trans}-azobenzene, the analysis implies an overrepresentation of certain structural characteristics following a photoexcitation at a given wavelength, potentially significant for the subsequent nuclear dynamics, photophysics, and photochemistry.
\end{abstract}
\maketitle

\section{Introduction}
A liquid system exhibits a wealth of local atomistic configurations, manifested as significant variation in {\it e.g.} its ensemble-averaged X-ray \cite{Wernet2004,Ottosson2011,Niskanen2016,Niskanen2017,Eronen2024} and UV--visible \cite{Timrov2016,Bononi2020,Chen2022} spectra. This underlying diversity is ignored when evaluating molecular properties for a single representative structure, and is probably impossible to capture by a few selected structures and their respective spectra \cite{Gomez2024}. Statistical variation of spectra implies pre-selection of structures for a spectral response, such as photon absorption at a given wavelength. At the same time, this structure-dependent response allows for structural characterization if the spectrally relevant structural degrees of freedom can be identified.
In a liquid system, this is not trivial. To this end, extensive simulations allow an insight into the structure--spectrum relationship when combined with analysis techniques such as the recently introduced emulator-based component analysis (ECA) \cite{Niskanen2022,Vladyka2023,Eronen2024a,Eronen2025}. This method is used to identify the characteristics of the input $\mathbf{x}$, that are decisive for the response $\mathbf{y}(\mathbf{x})$, thus simplifying the inverse problem posed by structural ($\mathbf{x}$) interpretation of spectra ($\mathbf{y}$). Such target-variance-driven dimensionality reduction is also useful when the outcome $\mathbf{y}$ depends predominantly on a small, unknown subset of features in a high-dimensional $\mathbf{x}$. The ECA method requires approximating $\mathbf{y}(\mathbf{x})$ by a reasonably accurate and computationally efficient emulator $\mathbf{y}_\mathrm{emu}(\mathbf{x})$, for which a neural network is a contemporary choice. Potential applications of the protocol include, but are not limited to, statistical phenomena in molecular liquids and in their spectra, the topic of this work.
\par
In this work, we study the UV--visible spectrum of {\it trans}-azobenzene in ethanol solution by combining molecular dynamics (MD) simulations with subsequent spectral calculations and dimensionality reduction procedures. We find that the dominant structural variation in this complex system, derived by principal component analysis (PCA), does not explain the intensity behavior within key spectral regions. Instead, decomposition by ECA captures the spectral variance with just a few latent structural degrees of freedom, that cover a small fraction of the total structural spread. With this low-dimensional representation, we identify the most significant structural characteristics behind a shift in the S$_2$ peak. Finally, we discuss the implications of these findings for processes occurring after photoexcitation.

\section{Methods}
We sampled 30\,000 structures from extended tight binding MD (xTBMD) using the GFN1-xTB method \cite{Grimme2017,Bannwarth2020} implemented in the CP2K software \cite{Kuhne2020} v.~2024.2. In the simulations, a {\it trans}-azobenzene molecule was dissolved in 159 ethanol molecules in a cubic box of $L$ = 25.04~{\AA} for the density 0.7840 g/cm$^3$ interpolated for pure ethanol liquid at 300 K \cite{Khattab2012}.
After equilibration of 20~ps, we sampled the canonical ensemble at 300~K with a 10~fs interval from a production run of 300~ps (for validation, see SI). We used a global Nos\'e--Hoover thermostat \cite{Nose1984,Nose1984b} with chain length 3, a time constant $\tau$ of 1000~fs, and an MD timestep of 0.5~fs. 
\par
For evaluation of spectra consisting of 50 transitions on each of the 30\,000 structures, we applied time-dependent density functional theory (TD-DFT) implemented in ORCA v.6.0.0 \cite{orca}. We explicitly included ethanol within the cutoff radius of 3.0~{\AA}, and simulated this cluster in a conductor-like polarizable continuum model \cite{GarciaRates2020}. We used B3LYP exchange-correlation potential \cite{Lee1988,Becke1993}, def2-SVP basis set \cite{Weigend2005} and def2/J auxiliary basis set \cite{Weigend2006} in the spectral simulations. We evaluated the cross section $\sigma_i$ of each transition $i$ for a randomly oriented sample using the transition matrix element in the velocity form. We then convolved this stick spectrum to differential cross section $\mathrm{d}\sigma/\mathrm{d}E$ in terms of energy $E$ by a Gaussian function (full-width at half maximum 0.414~eV), together with a constant shift of 0.300~eV in energy. We obtained these required parameters from a fit of the the computational ensemble-averaged spectrum to an experiment by Nägele and coworkers \cite{Nägele1997}. Finally, we transformed the differential cross section into wavelength $\lambda$ scale for $\mathrm{d}\sigma/{\mathrm{d}\lambda}$. We use a constant convolution width in energy, because a constant width in wavelength makes the lines artificially broad for short wavelengths.
\par

To capture the structure--spectrum relationship in these data, we train a neural network emulator and use it to iteratively find the ECA vectors for identifying spectrally decisive structural characteristics in the system. As the raw atomic coordinates from simulations are suboptimal for machine learning, we preprocess the structural information using a local version of the many-body tensor representation (LMBTR) \cite{Huo2022}. The LMBTR descriptor consists of Gaussian-smeared element-wise radial distance distributions for multiple centers in the system. As these centers, we use all the nitrogen and carbon sites in the azobenzene molecule, and eight additional virtual centers located perpendicular to a plane defined by nearby atoms in the azobenzene molecule. The first four virtual centers lie $\pm 2$~{\AA} from each nitrogen atom. The latter four centers lie $\pm 2$~{\AA} perpendicular from the arithmetic mean position of the {\it para} and {\it ortho} carbon atoms of each ring. For each structure, the encoding procedure resulted in a vector $\mathbf{d}$, which we z-score standardized to produce $\mathbf{\tilde{d}}$ used in machine learning and ECA. To establish a well-performing LMBTR--neural network combination, we ran a joint search \cite{Eronen2024a} including both LMBTR and neural network hyperparameters, and chose the best-performing combination. The final model had an R$^2$ score (coefficient of determination) of 0.945 on a test set of 5\,000 points, split from the data pool of 30\,000 points prior to model selection, and thus previously unseen by the model. For more details, we refer to SI.
\par
The ECA algorithm aims to identify a few spectrally most relevant structural degrees of freedom, ordered by their spectral significance. We project the z-score standardized structural feature vectors $\mathbf{\tilde{d}}$ of the test set onto the basis vectors $\mathbf{v}_i$. Thus, each basis vector results in a single latent coordinate $t_i$ for a data point. Furthermore, an approximation
\begin{equation}
\mathbf{\tilde{d}}^{(k)}=\sum_{i=1}^{k}t_i\mathbf{v}_i
\end{equation}
for each data point $\mathbf{\tilde{d}}$ is obtained in the given rank $k$. In the ECA procedure, vectors $\mathbf{v}_i$ are tuned rank-by-rank to maximize the explained spectral variance for the neural network emulator prediction over all data $\mathbf{\tilde{d}}^{(k)}$. We first optimize the basis vectors for dimensionality reduction using the training set, and then apply this dimensionality reduction in the analysis performed using the test set. The results of the analysis in the z-score standardized space can be converted to the original descriptor space by the corresponding inverse transform.
\begin{figure*}
    \centering
    \includegraphics[width=\textwidth]{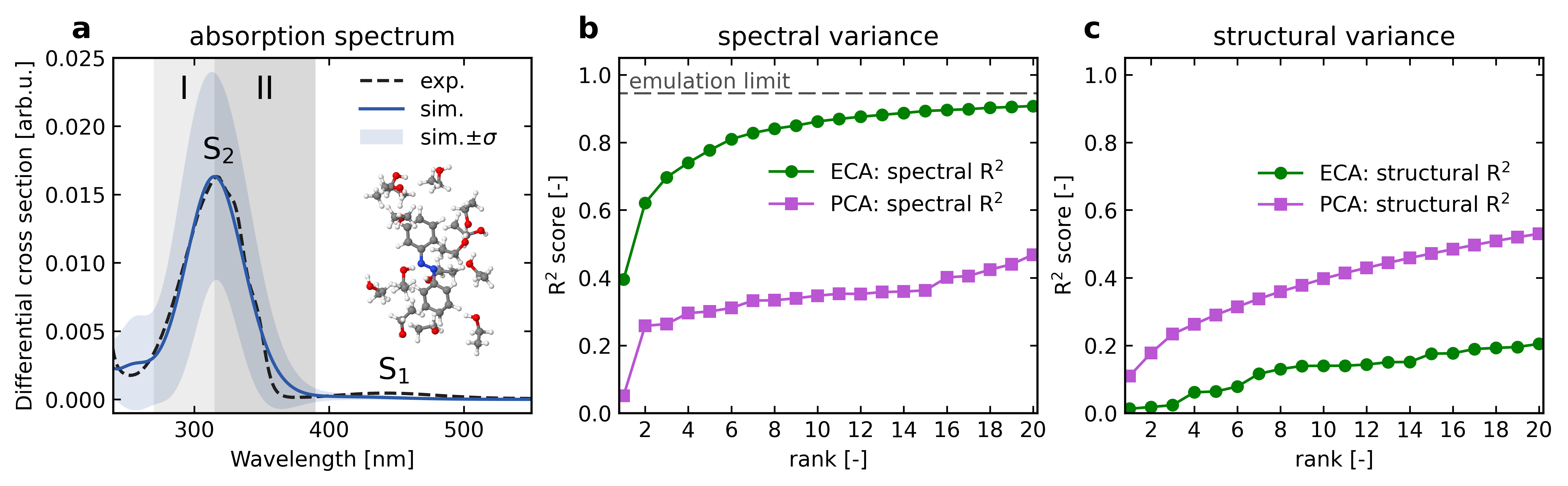}
    \caption{\label{fig:Spec_ECA} Results of the spectrum calculations and variance captured by different dimensionality reduction techniques. {\bf a}: Simulated ensemble average UV--visible absorption spectrum for {\it trans}-azobenzene in ethanol, together with an experiment reproduced from Ref.~\citenum{Nägele1997}. The blue-shaded area depicts the standard deviation of the data set, indicating significant statistical variation. The gray-shaded areas depict the two regions of interest (ROIs I and II) defined for the subsequent analysis. Computational results have been shifted to match the experiment. Additionally, the figure presents an example local structure prepared with Jmol \cite{jmol}. {\bf b}: Spectral R$^2$ score as a function of the rank of emulator-based component analysis (ECA) and principal component analysis (PCA), evaluated using the neural network emulator after the respective dimensionality reduction. The first two ECA components explain a large portion of the spectral ROI variance, while the higher components show rapidly diminishing R$^2$ score gains. Structural decomposition by PCA covers significantly less spectral variance for the same rank. {\bf c}: Structural R$^2$ score as a function of decomposition rank. The ECA vectors cover a small fraction of the structural variance compared to structural PCA. The result indicates that explaining structural variation alone does not explain the spectral variation.}
\end{figure*}

\section{Results}
Figure~\ref{fig:Spec_ECA}a shows the simulated ensemble-averaged spectrum together with the experimental one by Nägele and coworkers \cite{Nägele1997}. We apply the molecular notation S$_1$ and S$_2$ for the characteristic features in this spectrum, but note that in our simulation for the gas phase in the optimized geometry, the S$_2$ line already has three underlying states. In liquid, we find the photoexcited states of {\it trans}-azobenzene to be mixed with states of the ethanol, which results in a wealth of possible final states to be excited -- and to be simulated with explicit solvation. In our data, the average number of transitions within lines S$_1$ (390~nm--520~nm) and S$_2$ (270~nm--390~nm) were 0.9 and 14.1, respectively. The S$_1$ feature is almost unobservable, and we focus on the S$_2$ peak. We observe significant statistical variation between the spectra of individual local structures, as shown by the standard deviation $\sigma$ in Figure~\ref{fig:Spec_ECA}a. For the following analysis, we define two regions of interest (ROI), split at the maximum of the S$_2$ peak. Then, we study the mean differential cross section within each ROI as a function of the structure. This practice makes the analysis focus on overall features of the contributing spectra within the peak, instead of their fine details, that are more prone to error due to the used approximations. We first train a neural network to predict the two z-score standardized ROI values for a structure, and then utilize this emulator in in the ECA procedure for dimensionality reduction to identify the decisive structural characteristics behind the observed spectral variation. 
\par
The ECA method is analogous to PCA with the key difference being that the optimization target of the former is the explained ROI intensity variance. As seen in Figure \ref{fig:Spec_ECA}b, rapid convergence towards the emulator performance limit is observed when emulation is carried out on data points projected onto the low-dimensional subspace. In contrast, applying PCA to the structural data points $\mathbf{\tilde{d}}$ aims at finding the decisive components of structural variance, not necessarily reflected by the spectral response as seen in Figure~\ref{fig:Spec_ECA}b. Vectors from structural PCA cover significantly less spectral variance after the corresponding evaluation. Perhaps even more notably, the dominant structural degree of freedom is spectrally irrelevant. The different optimization goal of the methods is also evident in the structural R$^2$ score as a function of the decomposition rank, presented in Figure~\ref{fig:Spec_ECA}c, which shows PCA to surpass the ECA decomposition. Thus, the majority of the spectral ROI intensity variance is explained by a minority of structural variance. We note that another method, partial least squares fitting, has previously been compared to ECA, with the latter excelling in this comparison \cite{Niskanen2022}. 

Next, we turn to the analysis of the first two ECA component vectors, that indicate the structural changes responsible for most spectral ROI intensity variation. We first analyze the spectral effect, and then the {associated structural changes. We define two spectral properties: (i) ROI value difference (ROI I -- ROI II), and (ii) total ROI value sum. These properties can be directly linked in the two latent coordinates $t_1$ and $t_2$, as indicated in Figure~\ref{fig:ECAcorr}. The spectrally most significant structural degree of freedom $t_1$ corresponds to the ROI value difference (Figure~\ref{fig:ECAcorr}a), whereas the second most significant degree of freedom $t_2$ captures the ROI value sum (Figure~\ref{fig:ECAcorr}d). The two properties are mostly independent of each other, illustrated by the low cross-correlation in Figures~\ref{fig:ECAcorr}b and \ref{fig:ECAcorr}c.
From Figures~\ref{fig:ECAcorr}a and \ref{fig:ECAcorr}d we conclude that ROI I intensity is high when $t_1$ and $t_2$ values are high. Similarly, ROI II intensity is high when $t_1$ value is low and $t_2$ high. Thus, the structural changes causing a blueshift of the S$_2$ line are indicated by the latent coordinate $t_1$, as the effects from the second ECA component cancel out. For a structural interpretation of such effect, we calculate the difference 10\,$\mathbf{v}_1$ for a well-expected change in this coordinate, and then transform the respective ECA component vector to the original descriptor space. A further division by the squared distance to a corresponding center yields site-wise and element-wise curves with radial behavior proportional to that of the radial distribution function. 
\begin{figure}
    \centering
    \includegraphics[width=\columnwidth]{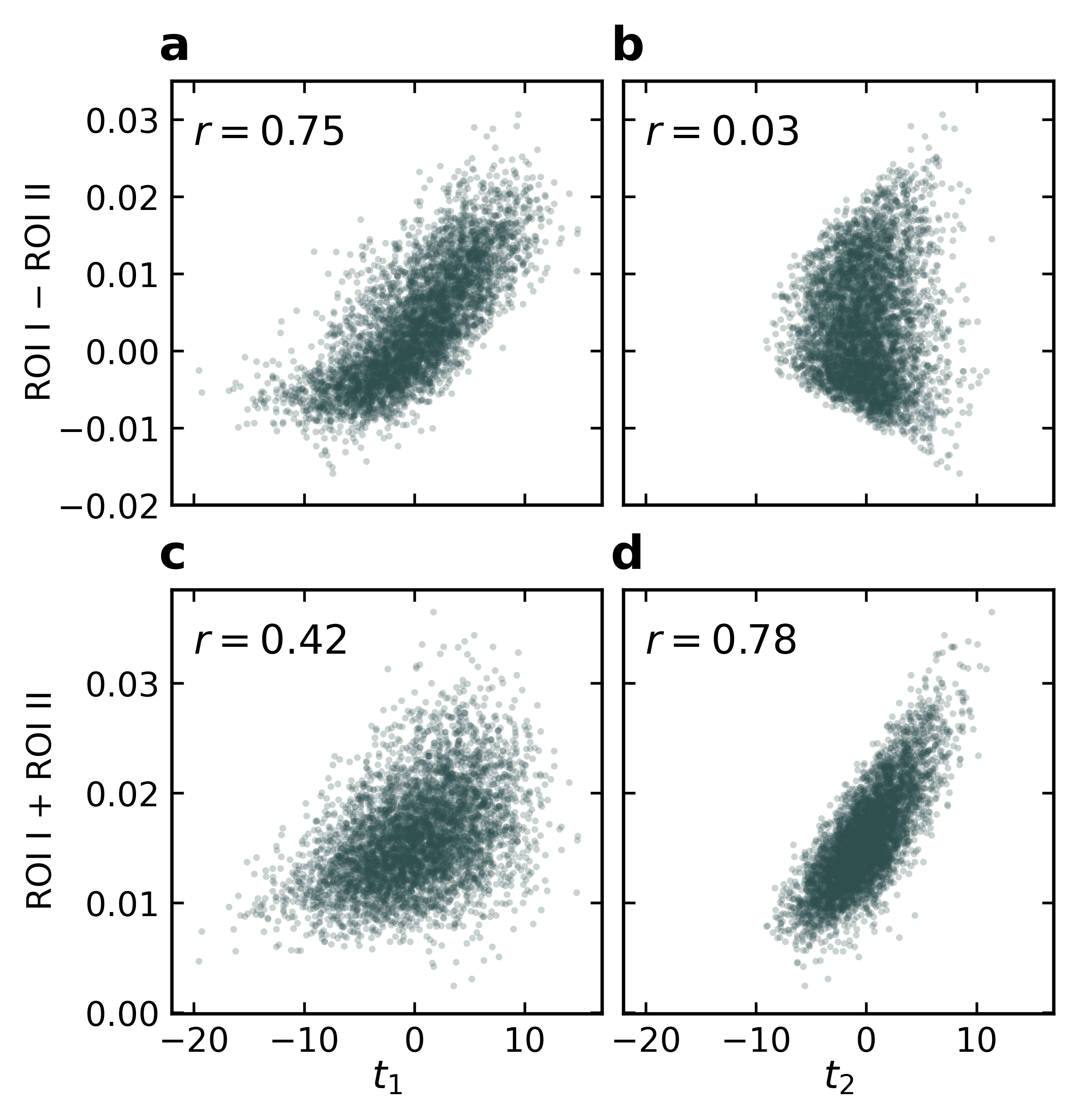}
    \caption{\label{fig:ECAcorr} The relationship between spectral properties and the latent coordinates from ECA using the test set with the corresponding Pearson's correlation coefficients $r$. {\bf a}: The difference between the two ROI values is highly correlated with the spectrally most significant structural variable $t_1$. {\bf b}: The correlation between ROI value difference and the second most significant structural variable $t_2$ is negligible. {\bf c}: The total ROI value sum is weakly correlated with $t_1$, but {\bf d}: highly correlated with $t_2$. For details, see text.}
\end{figure}

Figure~\ref{fig:ECAdiff} shows a selection of curves from the aforementioned comparison. The shorter-wavelength ROI I favors structures with less hydrogen bonding between the nitrogen atoms of {\it trans}-azobenzene and the OH group of the ethanol solvent. This is evident from the reduction of hydrogen and oxygen in the curves with an azobenzene N at the origin. This is accompanied by an increase of carbon near the nitrogen center, which we interpret as rotation of the ethanol molecule. Moreover, a blueshift is indicative of shorter N=N and C$_\mathrm{N=N}$--{\it ortho}-carbon bond lengths. The simultaneous reduction of N--C curves in short ranges indicates a shift towards longer N--C$_\mathrm{N=N}$ bond lengths at the azobenzene C--N=N--C bridge. The results were consistent across ten of the best-performing descriptor--neural network architecture combinations.

While the ECA procedure is carried out without {\it a priori} assumptions, the sanity of the obtained results can be confirmed {\it a posteriori} by evaluating the mean spectra for structures differing in the according characteristics. Using the whole data set, Figure \ref{fig:ECAconfirmed} shows a blueshift of the S$_2$ peak for structures with shorter N=N bond lengths and for structures with less hydrogen bonds for N atoms, evaluated using geometric criteria \cite{Niskanen2017}. We note that other notable structural changes are implied as well, and refer to the SI for the complete list. 

\begin{figure}
    \centering
    \includegraphics[width=\columnwidth]{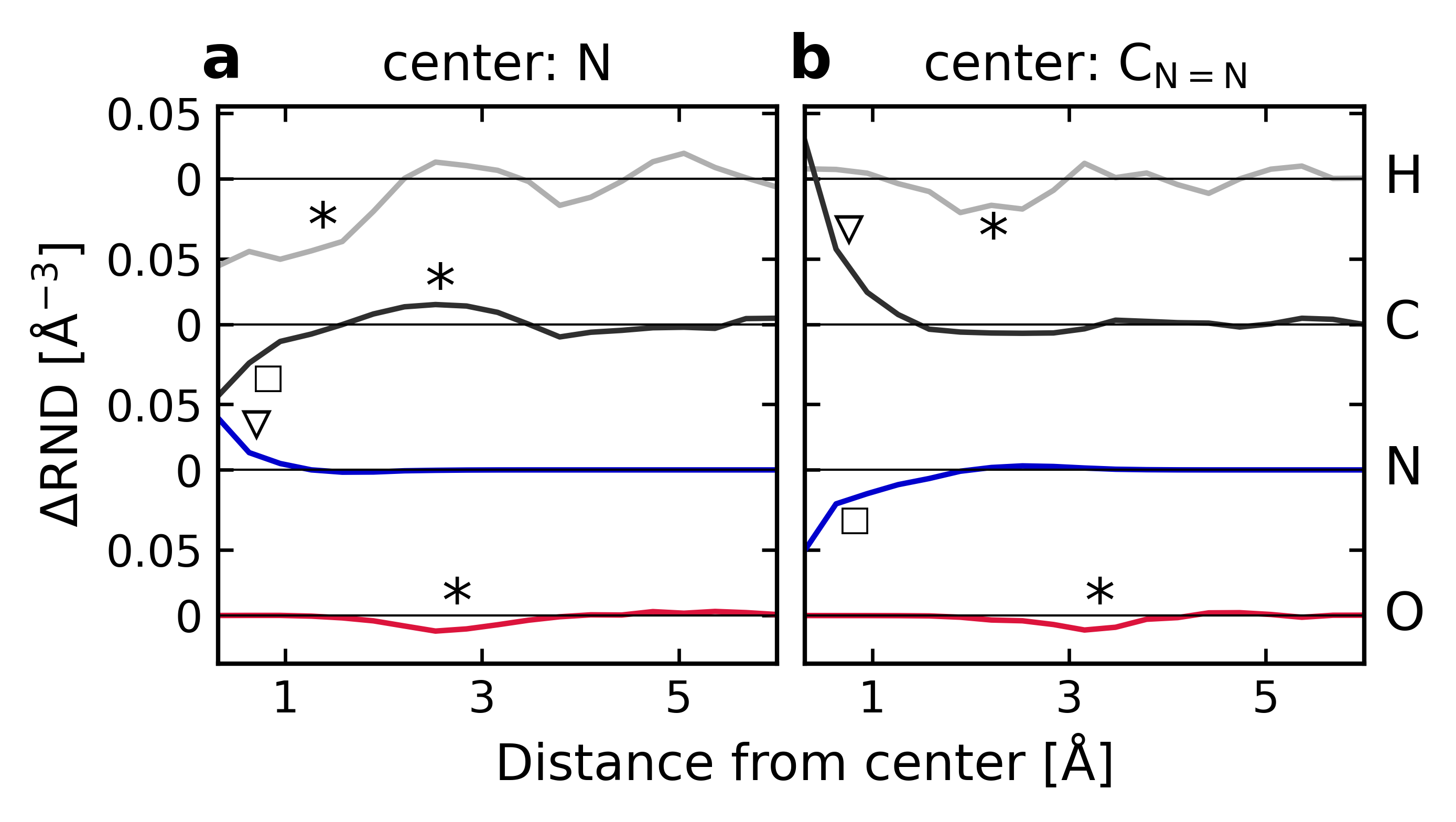}
    \caption{\label{fig:ECAdiff} Element-wise change in radial number density ($\Delta$RND) implied by a ten-unit rise of latent coordinate $t_1$ associated with ROI value difference from {\bf a}: the nitrogen atoms, and from {\bf b}: carbon atoms of the C-N=N-C bridge. We denote regions corresponding to the weakening of the hydrogen bonding and possible rotation of the ethanol molecules with an asterisk (*). Additionally, the triangles ($\triangledown$) denote regions indicating shortening of N=N or C$_\mathrm{N=N}$--C bonds, and the square ($\Box$) denotes a region indicating an increase in the length of the N--C bond. The curves shown are an average for the two N atoms or the two C$_\mathrm{N=N}$ atoms.}
\end{figure}
\begin{figure}
    \centering
    \includegraphics[width=\columnwidth]{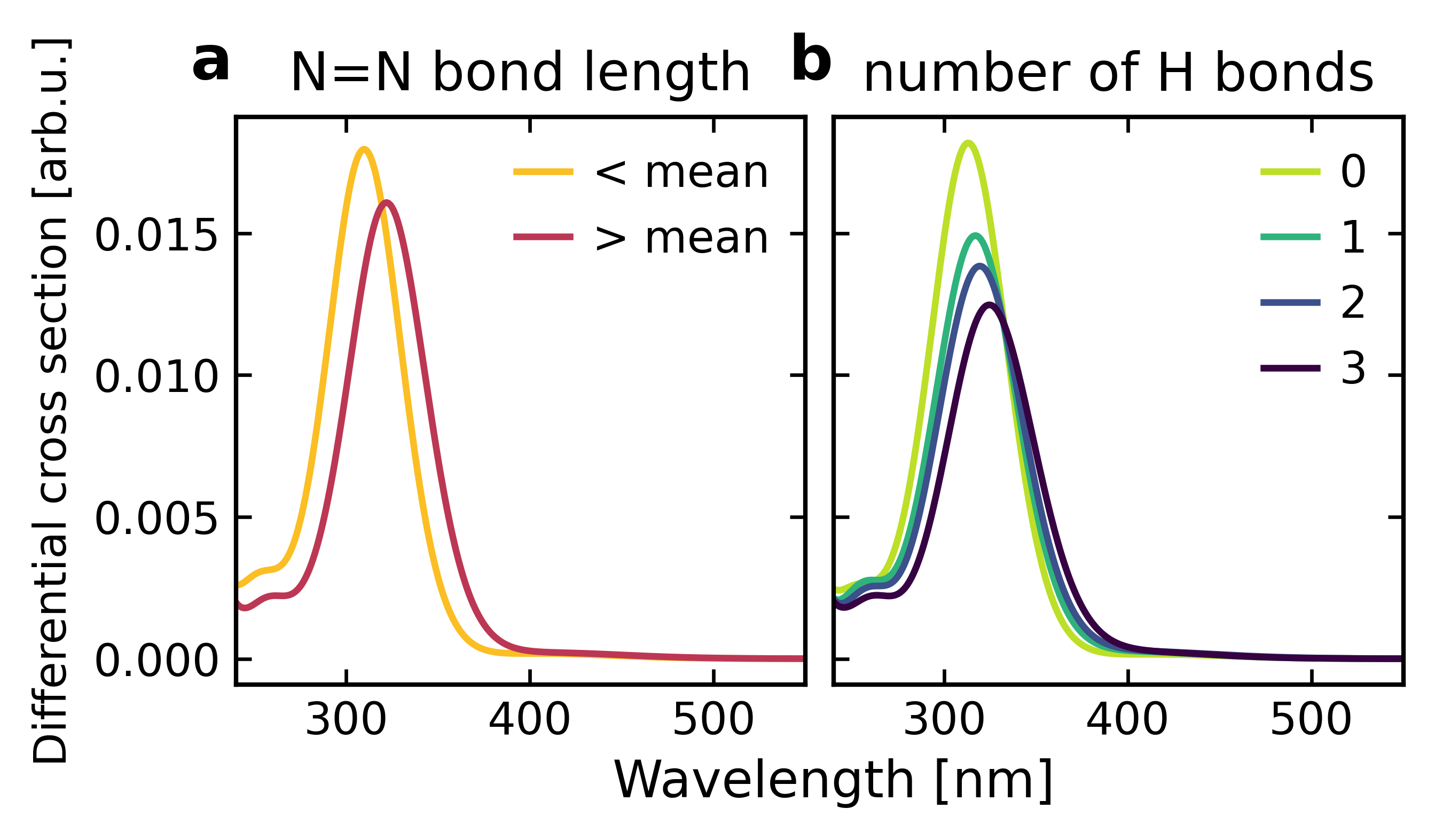}
    \caption{\label{fig:ECAconfirmed}Validation of the ECA results evaluated over the whole data set. {\bf a}: The mean spectra of structures with N=N bond length above and below the average value. {\bf b}: The mean spectra for structures with a given number of hydrogen bonds accepted by the azo group, calculated with criteria described in Ref. \cite{Niskanen2017}. The inspection is motivated by ECA first finding these characteristics.}
\end{figure}
\par

Additionally, we performed a corresponding analysis for the total ROI value sum, the results of which are presented in the SI. This spectral property appears to be affected, for example, increase in the amount of the ethanol solvent near the very tips of the carbon rings. The result can be affected by a molecule being excluded from the explicitly treated cluster due to the necessitated cutoff distance, but is still indicative of a closer ethanol molecule. The analysis also suggests that numerous other less interpretable structural features affect the ROI value sum.

We repeated the calculations using the PBE exchange–correlation potential \cite{pbe}, which in this case required evaluating 100 transitions. The aforementioned structural conclusions remain unaffected in the subsequent analysis, performed in a manner analogous to the presented one and reported in SI.

\section{Discussion}
The observed strong statistical variation of the UV--visible spectrum highlights the importance of proper ensemble sampling for the analysis of spectra of liquid systems. In this case, the variation is similar, if not greater, to what has been reported for X-ray spectra of liquids \cite{Ottosson2011,Niskanen2017,VazDaCruz2019,Eronen2024a,Eronen2025}. Furthermore, our analysis shows that the spectrum is affected by a wide range of structural properties, as a multitude of features in the ECA curves were non-zero (see Figure~\ref{fig:ECAdiff} and Supplementary Information). In spite of the large number of involved features, we emphasize that this subspace still covers a relatively small fraction of the total structural variance, as indicated in Figures~\ref{fig:Spec_ECA}b and \ref{fig:Spec_ECA}c. Last, the fact that the ECA curves show a contribution from oxygen atoms is a direct indication that the ethanol solvent has a significant effect on the spectrum.

During preliminary studies, we found that relatively large neural networks without regularization outperformed smaller or regularized ones. Interestingly, funnel-shaped neural networks with a linearly decreasing number of neurons at each layer seemed to allow for sufficient complexity while using less resources than their rectangle-shaped counterparts of uniform width. These large networks had a tendency to heavily overfit to the training data while still generalizing well to the validation data and, eventually, the test data. This phenomenon is likely an example of a machine learning model reaching the interpolating regime with emergent self-regularization capability \cite{Belkin2019}. In contrast, the model selection in our previous works with X-ray spectra favored simpler architectures \cite{Eronen2024,Eronen2024a,Eronen2025}. Furthermore, compared to the X-ray regime, a larger number of LMBTR centers with some virtual sites are necessary for sufficient machine learning performance for the UV--visible spectrum of ethanolic {\it trans}-azobenzene.  While the structure--spectrum relation in the X-ray regime is not necessarily straightforward, these results are a strong indication of an even greater complexity of UV--visible spectra. The conclusion is reasonable in the sense that both orbitals involved in the transition matrix element extend over a large volume, whereas the core-orbital participating in an X-ray process is quite localized.

The presented simulations necessitated numerous approximations. For example, the cutoff value for the explicit solvent is an obvious one. We chose the cutoff to be as large as possible while still enabling the simulation of $\sim$10$^4$ structure--spectrum pairs. Applying different methods for structural simulation and for property evaluation causes a bias in the results, but the practice is common in spectroscopy and quantum chemistry. In the case of this work, the MD run for the whole simulation cell would not be viable using the electronic structure method of the spectrum calculations. Additionally, we ignore the possible spectral effects arising from vibrational states. Despite this, the reasonability of our data is supported by the fact that the general shape of the spectrum agrees with the experiment. We also note that while switching from the B3LYP to the PBE exhange-correlation potential introduces more transitions (including possible ghost states) and slight changes in the spectra, the discussed structural dependencies of the ROI intensities are consistent across both models.

The challenges and limitations of the ECA method have been discussed in detail in Ref.~\citenum{Eronen2025}. Most notably, the interpretation of the ECA results is dependent on the properties of the descriptor, intended to encode all information required to predict a spectrum for a structure. While numerous alternatives exist, the LMBTR descriptor excelled in a comparison of six structural descriptor families for X-ray spectra \cite{Eronen2024a}, and the performance obtained in the current work for UV--visible aligns well with these earlier results \cite{Eronen2024,Eronen2024a,Eronen2025}. Further discussion on the required criteria for spectral interpretation by the protocol is available in Ref.~\citenum{Eronen2024}. In a nutshell, the descriptor must allow good machine learning and ECA performance, and also be interpretable by a human. Development of a new, highly interpretable, atomistic descriptor for machine learning can have a significant positive impact on the method in the future. 

Instead of representing two specific atomistic structures, the curves in Figure~\ref{fig:ECAdiff} reflect shifts in the structural distributions between the two selected points in the latent coordinate space. While not necessarily always the case, for the UV--visible spectrum of ethanolic {\it trans}-azobenzene, the two latent coordinates link the respective structural classes to a shift of the S$_2$ peak, and reveal potentially important implications for photoprocesses in liquids. Namely, as a consequence of the observed spectral behaviour, the used excitation wavelength introduces a preference for certain initial structures for the following photodynamics, potentially observable by, {\it e.g.}, pump--probe experiments. Solvent properties have also been found to have an effect on the photochemistry of azobenzene \cite{Bandara2012}, which could partially be a manifestation of this pre-selection.

\section{Conclusions}
The UV--visible absorption spectrum of ethanolic {\it trans}-azobenzene shows a broad statistical variation analogous to that found in X-ray spectra. Our subsequent analysis condensed the spectral variance to a few structural degrees of freedom, enabling a partial interpretation of the complex structure--spectrum relationship. We found that photon excitation on the short-wavelength side of the S$_2$ peak favors structures with, among other characteristics, weaker hydrogen bonding with the ethanol solvent, and a contraction of the N=N bond. Thus, the chosen incident photon energy may favor certain structures for the following photodynamics.

\section*{Data availability}
The data and relevant scripts are available in Zenodo: \href{https://doi.org/10.5281/zenodo.15349624}{10.5281/zenodo.15349624}.

\section*{Conflicts of interest}
There are no conflicts to declare.

\section*{Acknowledgements}
We acknowledge CSC – IT Center for Science, Finland, for providing computing resources. We thank Dr. N. Runeberg at CSC for support with the Puhti supercomputer, and Prof. J. Wachtveitl for providing the experimental UV--visible absorption spectrum of {\it trans}-azobenzene in ethanol solution. J.N. acknowledges funding by Research council of Finland via the academy project grant 367978.

\newpage
\bibliography{bibliography}

@Article{Weigend2005,
  author    = {Weigend, Florian and Ahlrichs, Reinhart},
  title     = {{Balanced basis sets of split valence{,} triple zeta valence and quadruple zeta valence quality for H to Rn: Design and assessment of accuracy}},
  journal   = {Physical Chemistry Chemical Physics},
  year      = {2005},
  volume    = {7},
  pages     = {3297-3305},
  doi       = {10.1039/B508541A},
  issue     = {18},
  publisher = {The Royal Society of Chemistry},
}

@Article{Weigend2006,
author ={Weigend, Florian},
title  ={{Accurate Coulomb-fitting basis sets for H to Rn}},
journal  ={Physical Chemistry Chemical Physics},
year  ={2006},
volume  ={8},
issue  ={9},
pages  ={1057-1065},
publisher  ={The Royal Society of Chemistry},
doi  ={10.1039/B515623H}
}

@Article{Bandara2012,
  author    = {Bandara, H. M. Dhammika and Burdette, Shawn C.},
  title     = {{Photoisomerization in different classes of azobenzene}},
  journal   = {Chemical Society Reviews},
  year      = {2012},
  volume    = {41},
  pages     = {1809-1825},
  doi       = {10.1039/C1CS15179G},
  issue     = {5},
  publisher = {The Royal Society of Chemistry},
}

@Article{Nägele1997,
  author   = {T. Nägele and R. Hoche and W. Zinth and J. Wachtveitl},
  title    = {{Femtosecond photoisomerization of cis-azobenzene}},
  journal  = {Chemical Physics Letters},
  year     = {1997},
  volume   = {272},
  number   = {5},
  pages    = {489-495},
  doi      = {10.1016/S0009-2614(97)00531-9},
}

@Article{Grimme2017,
author={Grimme, Stefan
and Bannwarth, Christoph
and Shushkov, Philip},
title={{A Robust and Accurate Tight-Binding Quantum Chemical Method for Structures, Vibrational Frequencies, and Noncovalent Interactions of Large Molecular Systems Parametrized for All spd-Block Elements (Z = 1--86)}},
journal={Journal of Chemical Theory and Computation},
year={2017},
publisher={American Chemical Society},
volume={13},
number={5},
pages={1989-2009},
doi={10.1021/acs.jctc.7b00118},
}

@Article{Bannwarth2020,
  author   = {Bannwarth, Christoph and Caldeweyher, Eike and Ehlert, Sebastian and Hansen, Andreas and Pracht, Philipp and Seibert, Jakob and Spicher, Sebastian and Grimme, Stefan},
  title    = {{Extended tight-binding quantum chemistry methods}},
  journal  = {WIREs Computational Molecular Science},
  year     = {2021},
  volume   = {11},
  number   = {2},
  pages    = {e1493},
  doi      = {10.1002/wcms.1493},
}

@Article{Kuhne2020,
  author   = {Kühne, Thomas D. and Iannuzzi, Marcella and Del Ben, Mauro and Rybkin, Vladimir V. and Seewald, Patrick and Stein, Frederick and Laino, Teodoro and Khaliullin, Rustam Z. and Schütt, Ole and Schiffmann, Florian and Golze, Dorothea and Wilhelm, Jan and Chulkov, Sergey and Bani-Hashemian, Mohammad Hossein and Weber, Valéry and Borštnik, Urban and Taillefumier, Mathieu and Jakobovits, Alice Shoshana and Lazzaro, Alfio and Pabst, Hans and Müller, Tiziano and Schade, Robert and Guidon, Manuel and Andermatt, Samuel and Holmberg, Nico and Schenter, Gregory K. and Hehn, Anna and Bussy, Augustin and Belleflamme, Fabian and Tabacchi, Gloria and Glöß, Andreas and Lass, Michael and Bethune, Iain and Mundy, Christopher J. and Plessl, Christian and Watkins, Matt and VandeVondele, Joost and Krack, Matthias and Hutter, Jürg},
  title    = {{CP2K: An electronic structure and molecular dynamics software package - Quickstep: Efficient and accurate electronic structure calculations}},
  journal  = {The Journal of Chemical Physics},
  year     = {2020},
  volume   = {152},
  number   = {19},
  pages    = {194103},
  doi      = {10.1063/5.0007045},
}

@Article{pbe,
  author  = {J. P. Perdew and K. Burke and M. Ernzerhof},
  title   = {{Generalized gradient approximation made simple}},
  journal = {Physical Review Letters},
  year    = {1996},
  volume  = {77},
  pages   = {3865--3868},
  doi     = {10.1103/PhysRevLett.77.3865},
}

@article{orca,
author = {Neese, Frank},
title = {{Software update: The ORCA program system—Version 5.0}},
journal = {WIREs Computational Molecular Science},
volume = {12},
number = {5},
pages = {e1606},
doi = {10.1002/wcms.1606},
year = {2022}
}

@Article{Khattab2012,
author={Khattab, Ibrahim Sadek
and Bandarkar, Farzana
and Fakhree, Mohammad Amin Abolghassemi
and Jouyban, Abolghasem},
title={{Density, viscosity, and surface tension of water+ethanol mixtures from 293 to 323K}},
journal={Korean Journal of Chemical Engineering},
year={2012},
volume={29},
number={6},
pages={812-817},
doi={10.1007/s11814-011-0239-6},
}

@Article{Huo2022,
  author    = {Haoyan Huo and Matthias Rupp},
  title     = {{Unified representation of molecules and crystals for machine learning}},
  journal   = {Machine Learning: Science and Technology},
  year      = {2022},
  volume    = {3},
  number    = {4},
  pages     = {045017},
  doi       = {10.1088/2632-2153/aca005},
  publisher = {IOP Publishing},
}

@Article{Niskanen2022,
  author  = {J. Niskanen and A. Vladyka and J. Niemi and C. J. Sahle},
  journal = {Royal Society Open Science},
  title   = {{Emulator-based decomposition for structural sensitivity of core-level spectra}},
  year    = {2022},
  volume  = {9},
  pages   = {220093},
  doi     = {10.1098/rsos.220093},
}

@article{Eronen2024,
doi = {10.1088/2399-6528/ad1f73},
year = {2024},
publisher = {IOP Publishing},
volume = {8},
number = {2},
pages = {025001},
author = {Eemeli A Eronen and Anton Vladyka and Florent Gerbon and Christoph J Sahle and Johannes Niskanen},
title = {{Information bottleneck in peptide conformation determination by x-ray absorption spectroscopy}},
journal = {Journal of Physics Communications},
}

@Article{Eronen2024a,
author ={Eronen, E. A. and Vladyka, A. and Sahle, Ch. J. and Niskanen, J.},
title  ={{Structural descriptors and information extraction from X-ray emission spectra: aqueous sulfuric acid}},
journal  ={Physical Chemistry Chemical Physics},
year  ={2024},
volume  ={26},
issue  ={34},
pages  ={22752-22761},
publisher  ={The Royal Society of Chemistry},
doi  ={10.1039/D4CP02454K},
}

@article{Eronen2025,
author = {Eronen, E. A. and Vladyka, A. and Sahle, Ch. J. and Niskanen, J.},
title = {{Structural Sensitivity of N 1s Excitations in N-Methylacetamide Solutions}},
journal = {The Journal of Physical Chemistry Letters},
volume = {16},
number = {7},
pages = {1666-1672},
year = {2025},
doi = {10.1021/acs.jpclett.4c03487}
}

@Article{Belkin2019,
  author   = {Mikhail Belkin and Daniel Hsu and Siyuan Ma and Soumik Mandal},
  title    = {{Reconciling modern machine-learning practice and the classical bias–variance trade-off}},
  journal  = {Proceedings of the National Academy of Sciences},
  year     = {2019},
  volume   = {116},
  number   = {32},
  pages    = {15849-15854},
  doi      = {10.1073/pnas.1903070116}
}

@Article{Wernet2004,
  author  = {Ph. Wernet and D. Nordlund and U. Bergmann and M. Cavalleri and M. Odelius and H. Ogasawara and L. {\AA}. N\"aslund and T. K. Hirsch and L. Ojam\"ae and P. Glatzel and L. G. M. Pettersson and A. Nilsson},
  title   = {{The Structure of the First Coordination Shell in Liquid Water}},
  journal = {Science},
  year    = {2004},
  volume  = {304},
  number  = {5673},
  pages   = {995-999},
  doi     = {10.1126/science.1096205},
}

@Article{Ottosson2011,
  author    = {Ottosson, Niklas and Børve, Knut J. and Spångberg, Daniel and Bergersen, Henrik and Sæthre, Leif J. and Faubel, Manfred and Pokapanich, Wandared and Öhrwall, Gunnar and Björneholm, Olle and Winter, Bernd},
  journal   = {Journal of the American Chemical Society},
  title     = {{On the Origins of Core--Electron Chemical Shifts of Small Biomolecules in Aqueous Solution: Insights from Photoemission and ab Initio Calculations of Glycineaq}},
  year      = {2011},
  number    = {9},
  pages     = {3120--3130},
  volume    = {133},
  doi       = {10.1021/ja110321q},
  publisher = {American Chemical Society (ACS)},
}

@Article{Niskanen2017,
  author  = {Niskanen, Johannes and Sahle, Christoph J. and Gilmore, Keith and Uhlig, Frank and Smiatek, Jens and F{\"o}hlisch, Alexander},
  title   = {{Disentangling Structural Information From Core-level Excitation Spectra}},
  journal = {Physical Review E},
  year    = {2017},
  volume  = {96},
  pages   = {013319},
  doi     = {10.1103/PhysRevE.96.013319},
}

@misc{jmol,
  author = {{Jmol development team}},
  title = {{Jmol: an open-source Java viewer for chemical structures in 3D}},
  url = {http://www.jmol.org/},
  urldate = {2025-01-21},
  note = {Accessed: January 21, 2025}
}

@Article{VazDaCruz2019,
  author  = {Vaz da Cruz, V. and Gel'mukhanov, F. and Eckert, S. and Iannuzzi, M. and Ertan, E. and Pietzsch, A. and Couto, R. C. and Niskanen, J. and Fondell, M. and Dantz, M. and Schmitt, T. and Lu, X and McNally, D. and Jay, R. M. and Kimberg, V. and F\"ohlisch, A. and Odelius, M.},
  title   = {{Probing Hydrogen Bond Strength in Liquid Water by Resonant inelastic X-ray scattering}},
  journal = {Nature Communications},
  year    = {2019},
  volume  = {10},
  pages   = {1013},
  doi     = {10.1038/s41467-019-08979-4}
}

@article{Timrov2016,
author = {Timrov, Iurii and Micciarelli, Marco and Rosa, Marta and Calzolari, Arrigo and Baroni, Stefano},
title = {{Multimodel Approach to the Optical Properties of Molecular Dyes in Solution}},
journal = {Journal of Chemical Theory and Computation},
volume = {12},
number = {9},
pages = {4423-4429},
year = {2016},
doi = {10.1021/acs.jctc.6b00417}
}

@article{Bononi2020,
author = {Bononi, Fernanda C. and Chen, Zekun and Rocca, Dario and Andreussi, Oliviero and Hullar, Ted and Anastasio, Cort and Donadio, Davide},
title = {{Bathochromic Shift in the UV–Visible Absorption Spectra of Phenols at Ice Surfaces: Insights from First-Principles Calculations}},
journal = {The Journal of Physical Chemistry A},
volume = {124},
number = {44},
pages = {9288-9298},
year = {2020},
doi = {10.1021/acs.jpca.0c07038}
}

@article{Chen2022,
author = {Chen, Zekun and Bononi, Fernanda C. and Sievers, Charles A. and Kong, Wang-Yeuk and Donadio, Davide},
title = {{UV–Visible Absorption Spectra of Solvated Molecules by Quantum Chemical Machine Learning}},
journal = {Journal of Chemical Theory and Computation},
volume = {18},
number = {8},
pages = {4891-4902},
year = {2022},
doi = {10.1021/acs.jctc.1c01181},
}

@Article{Gomez2024,
AUTHOR = {Gómez, Sara and Lafiosca, Piero and Giovannini, Tommaso},
TITLE = {{Modeling UV/Vis Absorption Spectra of Food Colorants in Solution: Anthocyanins and Curcumin as Case Studies}},
JOURNAL = {Molecules},
VOLUME = {29},
YEAR = {2024},
NUMBER = {18},
pages = {4378},
DOI = {10.3390/molecules29184378}
}

@Article{GarciaRates2020,
  author   = {Garcia-Ratés, Miquel and Neese, Frank},
  title    = {{Effect of the Solute Cavity on the Solvation Energy and its Derivatives within the Framework of the Gaussian Charge Scheme}},
  journal  = {Journal of Computational Chemistry},
  year     = {2020},
  volume   = {41},
  number   = {9},
  pages    = {922-939},
  doi      = {10.1002/jcc.26139},
}

@Article{Vladyka2023,
  author    = {Vladyka, Anton and Sahle, Christoph J. and Niskanen, Johannes},
  title     = {{Towards structural reconstruction from X-ray spectra}},
  journal   = {Physical Chemistry Chemical Physics},
  year      = {2023},
  volume    = {25},
  number    = {9},
  pages     = {6707-6713},
  doi       = {10.1039/D2CP05420E},
  issue     = {9},
  publisher = {The Royal Society of Chemistry},
}

@Article{Niskanen2016,
  author   = {Niskanen, J. and Sahle, Ch. J. and Ruotsalainen, K. O. and M\"uller, H. and Kav{\v c}i{\v c}, M. and {\v Z}itnik, M. and Bu{\v c}ar, K. and Petric, M. and Hakala, M. and Huotari, S.},
  title    = {{Sulphur K$\beta$ emission spectra reveal protonation states of aqueous sulfuric acid}},
  journal  = {Scientific Reports},
  year     = {2016},
  volume   = {6},
  pages    = {21012},
  doi = {10.1038/srep21012}
}

@article{Nose1984,
author  = {S. Nose}, 
title   = {{A unified formulation of the constant temperature molecular dynamics methods}},
journal = {The Journal of Chemical Physics}, 
volume  = {81}, 
pages   = {511--519},
year    = {1984},
doi     = {10.1063/1.447334}
}

@article{Nose1984b,
    author  = {S. Nose}, 
    title   = {{A molecular dynamics method for simulations in the canonical ensemble}}, 
    journal = {Molecular Physics},
    volume  = {52}, 
    pages   = {255--268},
    year    = {1984},
    doi     = {10.1080/00268978400101201}
}

@article{Lee1988,
  title = {{Development of the Colle-Salvetti correlation-energy formula into a functional of the electron density}},
  author = {Lee, Chengteh and Yang, Weitao and Parr, Robert G.},
  journal = {Phys. Rev. B},
  volume = {37},
  issue = {2},
  pages = {785--789},
  year = {1988},
  publisher = {American Physical Society},
  doi = {10.1103/PhysRevB.37.785},
}

@article{Becke1993,
    author = {Becke, Axel D.},
    title = {{Density‐functional thermochemistry. III. The role of exact exchange}},
    journal = {The Journal of Chemical Physics},
    volume = {98},
    number = {7},
    pages = {5648-5652},
    year = {1993},
    doi = {10.1063/1.464913},
}

%

\onecolumngrid
\clearpage
\section*{Supplementary Information}

\subsection{Validation of the molecular dynamics calculations}

To validate the sampling of the molecular dynamics simulations, we evaluated autocorrelation functions of the production run sampled at 10~fs intervals. Figure~\ref{fig:autocorr} shows the autocorrelation for ROI intensities, LMBTR vectors, and full spectra. The curves for the ROI values and the full spectra show a rapid initial drop, indicating low correlation even for two temporally subsequent sampled data points. Therefore, the structure--spectrum data set used in this work consists of mostly uncorrelated data points even if the structural autocorrelation decays slower than its spectral counterpart.

\begin{figure}[h!]
    \centering
    \includegraphics[width=0.7\linewidth]{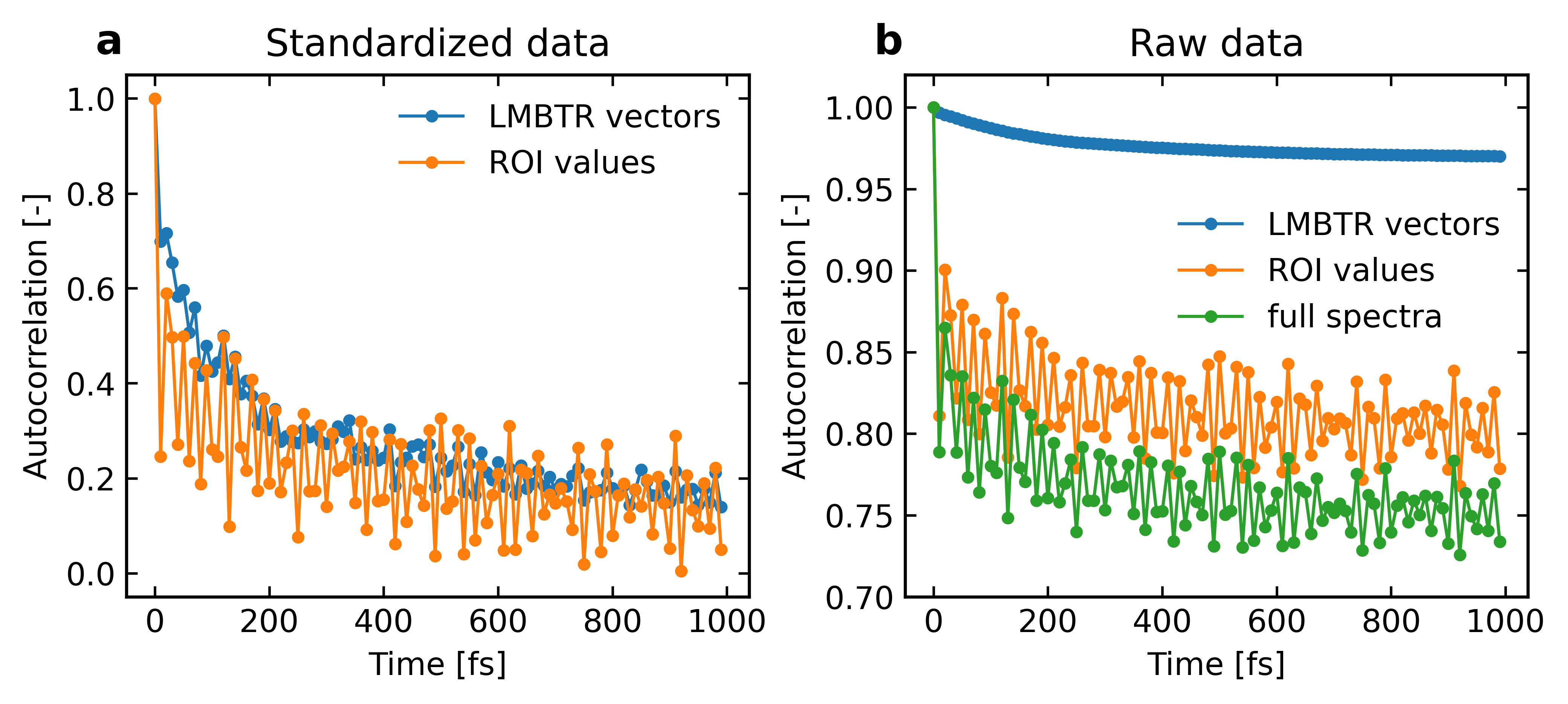}
    \caption{Autocorrelation functions of the production run at intervals of 10~fs. {\bf a}: The curves for z-score standardized data which was used in the analysis. Spectral data is represented by the two regions of interest values (ROIs) and the structural data is represented by the local many-body tensor representation (LMBTR) vectors derived from the xyz-coordinates and atomic number of each nucleus. {\bf b} Corresponding curves for data that has not been z-score standardized. Additionally, we show the autocorrelation of the full spectra.}
    \label{fig:autocorr}
\end{figure}

\subsection{Machine learning details}

We split the data set of 30~000 structure--spectrum pairs into three parts: (i) a training set of 24~000 points for training the neural network and fitting the ECA vectors, (ii) a validation set of 1~000 points for early stopping of neural network training and (iii) a test set of 5~000 points for testing the model and ECA. For the performance metric we chose the coefficient of determination or R$^2$ score which represents the spectral variance covered by the model. An R$^2$ score of 1 refers to perfect prediction performance whereas a value of 0 refers to performance where the model always predicts the mean of the data set. We z-score standardized the LMBTR vectors and the ROI values before any machine learning tasks. This means that we subtracted the according training data mean from each of the vector components, after which we divided the values by their standard deviations in the training set. 

As the machine learning model we chose a fully connected feed-forward neural network emulator implemented using PyTorch\footnote{A. Paszke {\it et. al.} PyTorch: An Imperative Style, High-Performance Deep Learning Library. Advances in Neural Information Processing Systems 32, 8024–8035, 2019. url:https://dl.acm.org/doi/10.5555/3454287.3455008{.}}. As illustrated in Figure~\ref{fig:NN}, we used funnel-shaped neural networks with linearly decreasing number of neurons at each layer (excluding the input layer). Both the descriptor and the neural network have a wide range of tunable hyperparameters. Therefore, we ran a joint hyperparameter search\footnote{E. A. Eronen, A. Vladyka, Ch. J. Sahle, and J. Niskanen. Structural descriptors and information extraction from X-ray emission spectra: aqueous sulfuric acid. Physical Chemistry Chemical Physics, 26:22752–22761, 2024. doi:10.1039/D4CP02454K.} including both the descriptor and neural network hyperparameters as presented in Table~\ref{tab:MS}. We trained the models to predict the z-score standardized ROI intensities in mini-batches of 512 points each. The training was stopped if the performance on the separate early stopping data set of 1\,000 points did not improve in 10 consecutive epochs. We used the rectified linear unit activation function and the Adam optimizer with an initial learning rate of 10$^{-3}$ to minimize the R$^2$ loss (equals $1 - \mathrm{R}^2 $ score). 

For selection of the best combination of hyperparameters, we used five-fold cross-validation using the aforementioned training method to evaluate the performance of each trial model. In the end, we chose the model with the best mean cross-validation R$^2$ score and trained the final model using the full training data of 24000 data points. 

For ECA we used the implementation by Vladyka {\it et al.}\footnote{A. Vladyka, E. A. Eronen, and J. Niskanen, Implementation of the emulator-based component analysis, Journal of Computational Science 83, 102437 (2024). doi:10.1016/j.jocs.2024.102437}. To ensure that a proper minimum is found, we fitted the each ECA vector 25 times with different initial guesses and chose the one with best performance before proceeding with the optimization of the next vector.

\begin{figure}[h!]
    \centering
    \includegraphics[width=0.33\columnwidth]{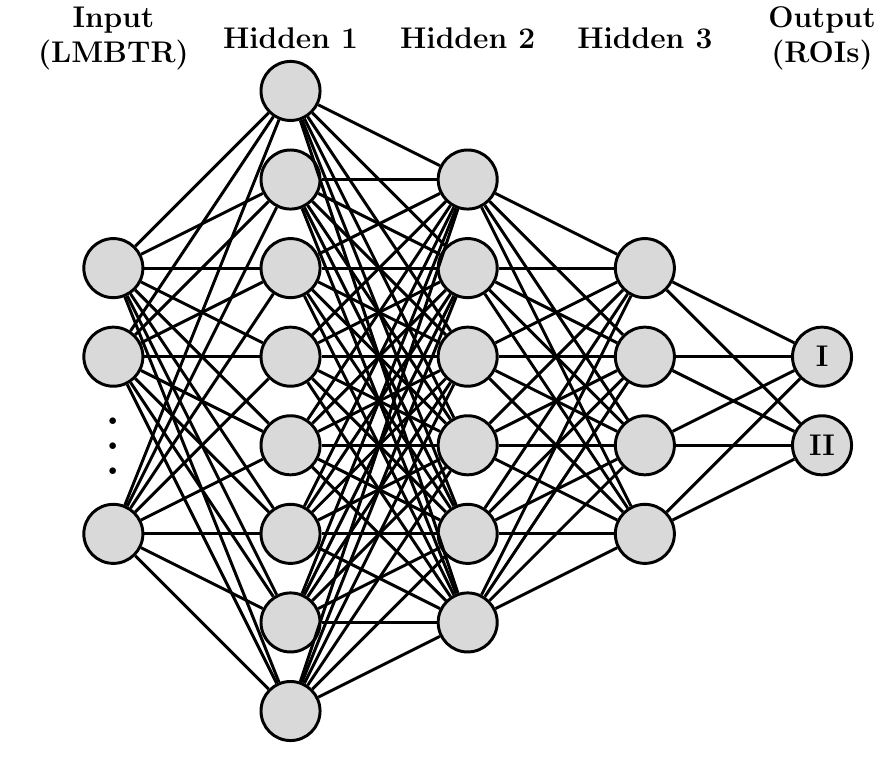}
    \caption{\label{fig:NN} A schematic of a funnel-shaped neural network used in this work. }
\end{figure}

\begin{table*}[h!]
\centering
\caption{The grid for the joint search of the LMBTR and neural network (NN) hyperparameters. The best performing combination is indicated in bold face.} \label{tab:MS}
\begin{tabular}{l|l|l}\label{tab:emulator_scores}
 & Hyperparameter & Grid points \\
\hline
LMBTR  & grid min, \AA & \{{\bf 0}\} \\
       & grid max, \AA & \{4, 5, {\bf 6}, 7\} \\
       & number of grid points & \{{\bf 20}, 40\} \\
       & Gaussian width, \AA & \{0.4, {\bf 0.6}, 0.8, 1.0\} \\
NN     & weight decay &\{{\bf 0}\} \\
       & number of hidden layers & \{2, {\bf 3}, 4, 5, 6\}\\
       & width of the first hidden layer & \{1024, {\bf 2048}, 4096, 8192, 16384\} \\
\end{tabular}
\end{table*}

\newpage
\subsection{A schematic of all LMBTR centers}

Figure~\ref{fig:molecule} shows a schematic of all the centers that we included in the LMBTR descriptor. The notation is used throughout the analysis of the structure--spectrum relationship. In the main text, we show the average of element-wise curves centered at N0 and N1, and the average of element-wise curves centered at C5 and C8.

\begin{figure}[h!]
    \includegraphics[width=0.5\columnwidth]{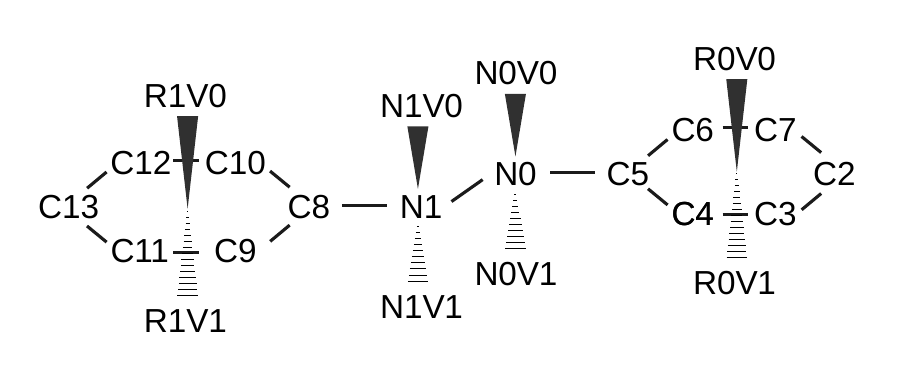}
    \caption{\label{fig:molecule} A schematic of the centers for the LMBTR descriptor of the azobenzene molecule. We chose all the nitrogen and carbon atoms of the molecule and a set of eight virtual points (V) as centers. The first four are located  $\pm$2~{\AA} off-plane from the nitrogen atoms, with the plane defined by three nearby atoms (N1, N0 and C5, or N0, N1 and C8). The latter four are located $\pm$2~{\AA} off-plane from the carbon rings (R), with the plane defined by the two {\it ortho}-carbon atoms (C4 and C6, or C9 and C10) and a {\it para}-carbon atom (C2 or C13).}
\end{figure}

\newpage
\subsection{ROI value difference with large overall intensity: full list of structural changes}

Figure~\ref{fig:ROIdiff_full} shows the complete list of structural changes implied by a blueshift in the S$_2$ peak, according to the applied analysis procedure. The vector, originally in the z-score-standardized descriptor space, has been inverse transformed to the original descriptor space after which each feature has been divided by the squared distance to the corresponding center. In addition to the results discussed in the main text, the curves at centers C2 and C13 indicate shortening of the {\it para}-carbon--{\it meta}-carbon bonds. Other less interpretable effects are also implied by the curves; for example, the hydrogen and carbon curves of the virtual centers. We note that the large absolute values near zero exhibited by some of the curves of the virtual centers are an artifact originating from the division by the squared distance to the center. 
\begin{figure}[h!]
    \centering
    \includegraphics[width=0.8\columnwidth]{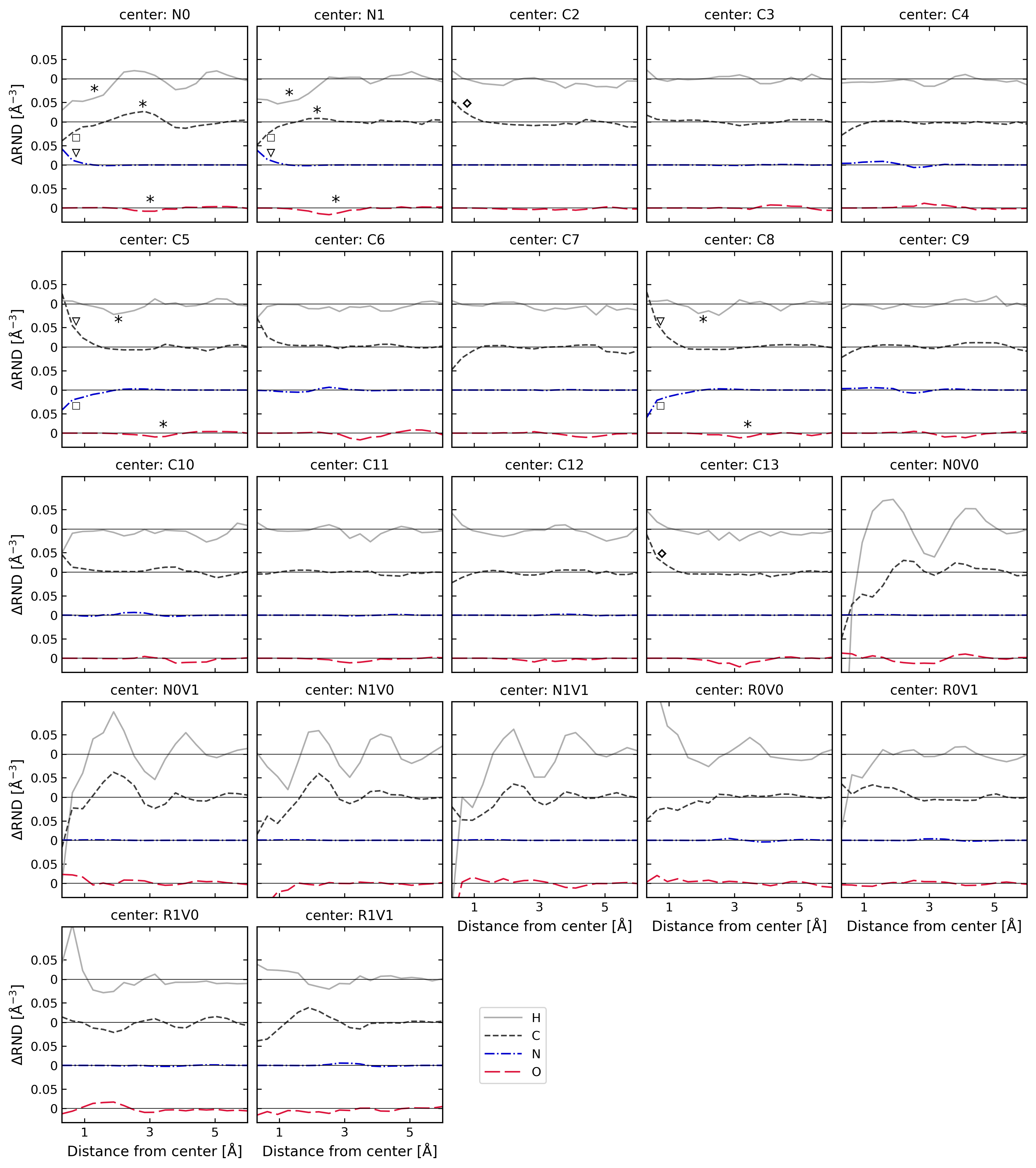}
    \caption{\label{fig:ROIdiff_full} The full LMBTR vector corresponding to the structural shift behind the increase of the ROI value difference. We denote regions corresponding to weakening of hydrogen bonds with an asterisk (*), shortening of N=N bond with triangles ($\triangledown$), shortening of C--C bonds with diamonds ($\diamond$), and increase in the N--C bond lengths with squares ($\Box$). Numerous other less interpretable factors also play a role in the structure--spectrum relationship.}
\end{figure}

\newpage
\subsection{Structural changes implied by the increase of total ROI value}

Similarly to the analysis of the ROI value difference, Figure~\ref{fig:ROIsum_full} shows the complete list of structural changes implied by an increase in the total ROI value, {\it i.e.} the sum of the average differential cross section within each ROI. Again, several structural factors affect the spectrum, some being more interpretable than others. The ROI sum is affected by, for example, an increase in the amount of the ethanol solvent near the very tips of the carbon rings. The result can be affected by a molecule being excluded from the explicitly treated cluster due to the necessitated cutoff distance, but is still indicative of a closer ethanol molecule.

\begin{figure}[h!]
    \centering
    \includegraphics[width=0.8\columnwidth]{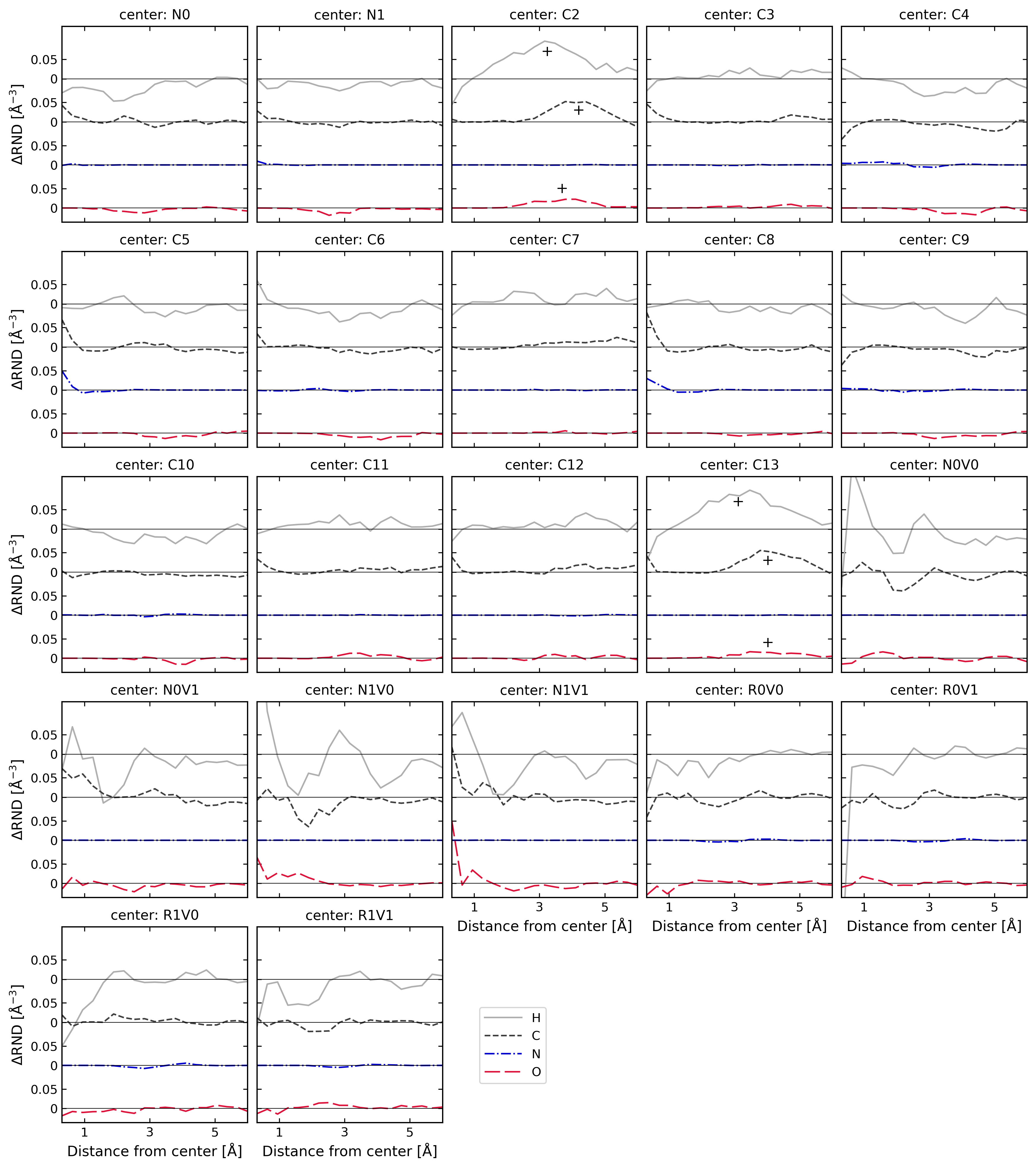}
    \caption{\label{fig:ROIsum_full} The full LMBTR vector corresponding to the structural shift behind the increase in total ROI value. We denote regions corresponding to the increase in the amount of ethanol solvent near the tips of the azobenzene rings with plus signs (+). Many other less interpretable factors also play a role in the structure--spectrum relationship.}
\end{figure}

\subsection{Results using spectra calculated with the PBE exchange-correlation potential}

In the main text we studied regions of interest derived from spectra calculated using the B3LYP functional. Here we present the corresponding results using spectra calculated with the PBE exchange-correlation potential. Figure~\ref{fig:F1PBE} shows the ensemble averaged spectrum, where each line of the individual spectra is convolved with a Gaussian of full-width at half maximum of 0.370~eV and shifted by 0.527~eV. Analogous to the case of B3LYP in the main text, we obtained these values by a fitting the ensemble-averaged spectrum on the experiment by Nägele and coworkers\footnote{T. Nägele {\it et. al.}: Femtosecond photoisomerization of cis-azobenzene. Chemical Physics Letters 272, 489, 1997. doi:10.1016/S0009-2614(97)00531-9}. Especially at the low-wavelength tail of the S$_2$ peak, the simulated spectrum approximates the experiment less well than the corresponding spectrum calculated with B3LYP. The R$^2$ score behavior of the ECA decomposition is similar between the two cases, with the spectral R$^2$ score rising slightly faster with PBE. Figure~\ref{fig:F2PBE} shows that the first two ECA components show the similar behavior with either of the functionals, except the order of the vectors one and two are flipped. Figure~\ref{fig:F3PBE} shows that curves denoting structural changes behind a blueshift in the spectrum show highly similar behavior to that found in the main text. Therefore, the qualitative conclusions of the main text are the same for PBE and B3LYP. The different behavior at short distances is likely due to the different LMBTR Gaussian widths originating from the independent model selections performed for B3LYP and PBE data. Figure~\ref{fig:F4PBE} confirms the consistency in terms of N=N bond lengths and the number of hydrogen bonds.

\begin{figure}[h!]
    \centering
    \includegraphics[width=1\columnwidth]{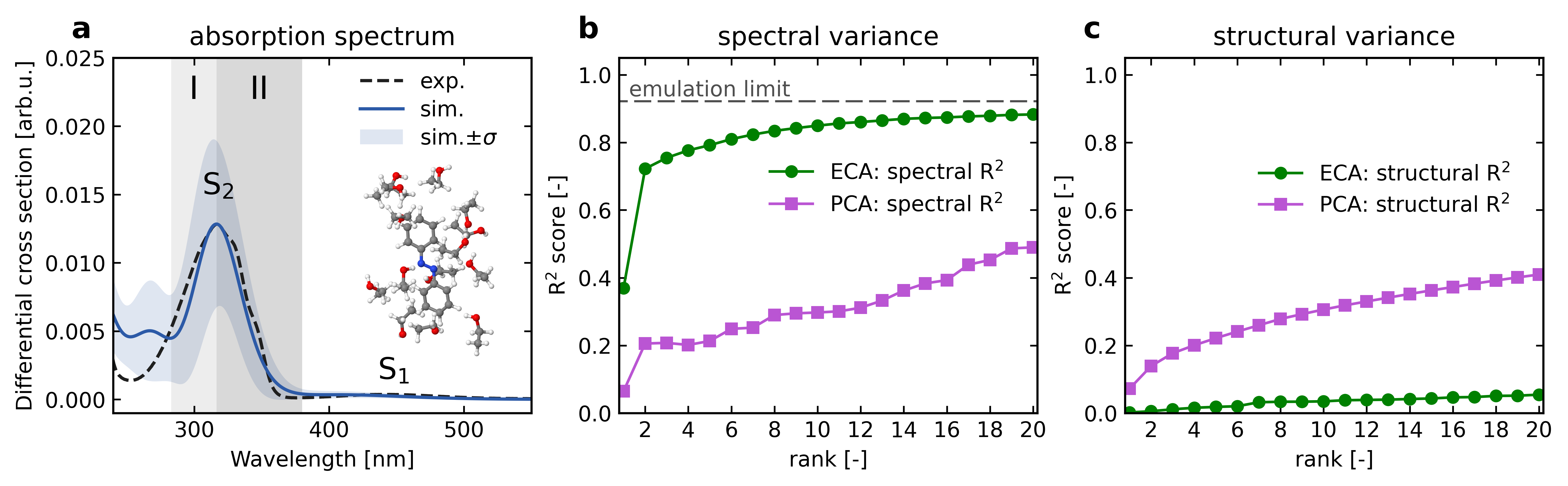}
    \caption{\label{fig:F1PBE} The corresponding presentation of Figure~1 of the main text, obtained using the PBE functional. }
\end{figure}

\begin{figure}[h!]
    \centering
    \includegraphics[width=0.5\columnwidth]{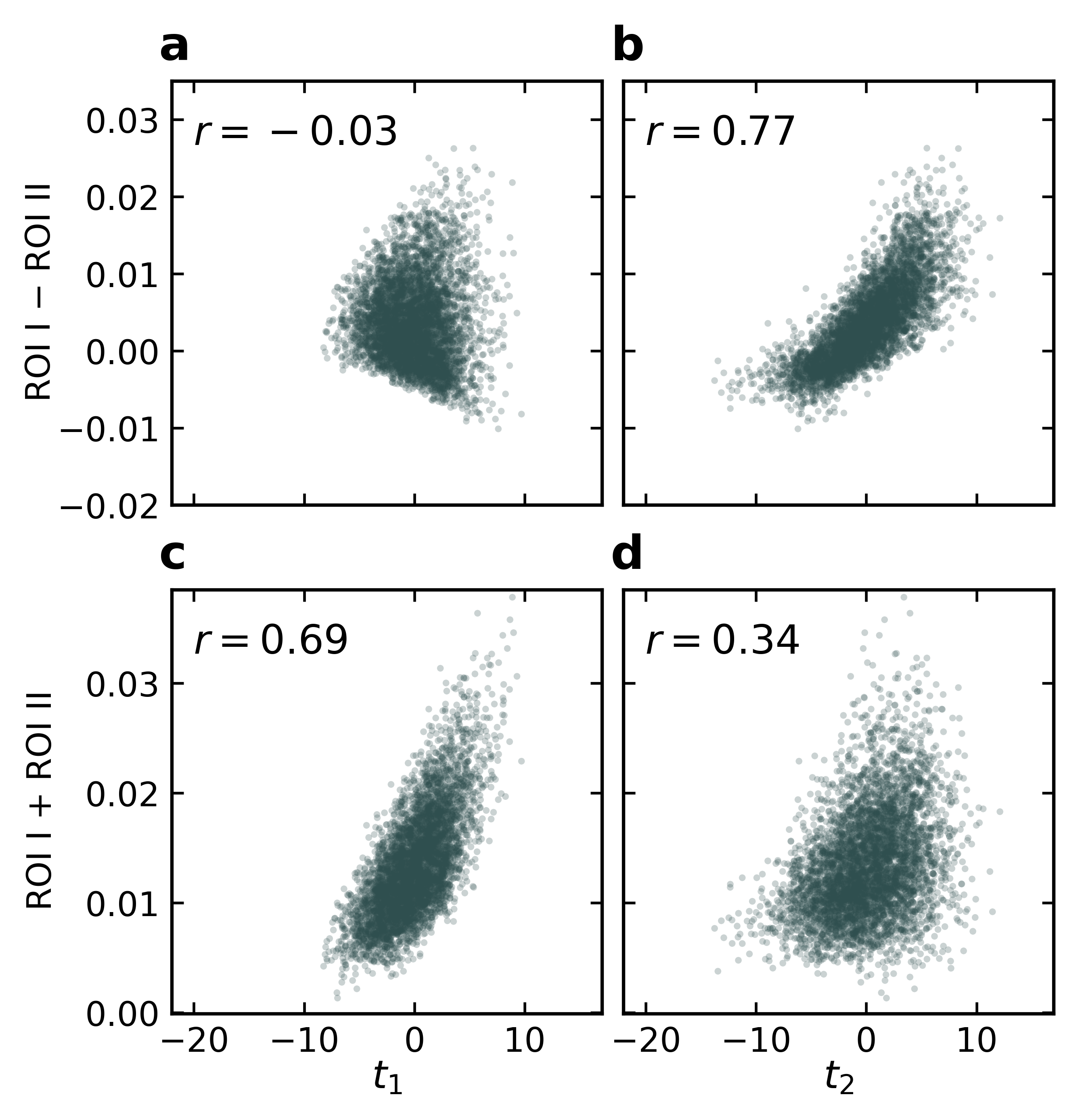}
    \caption{\label{fig:F2PBE} The corresponding presentation of Figure~2 of the main text, obtained using the PBE functional.}
\end{figure}

\begin{figure}[h!]
    \centering
    \includegraphics[width=0.5\columnwidth]{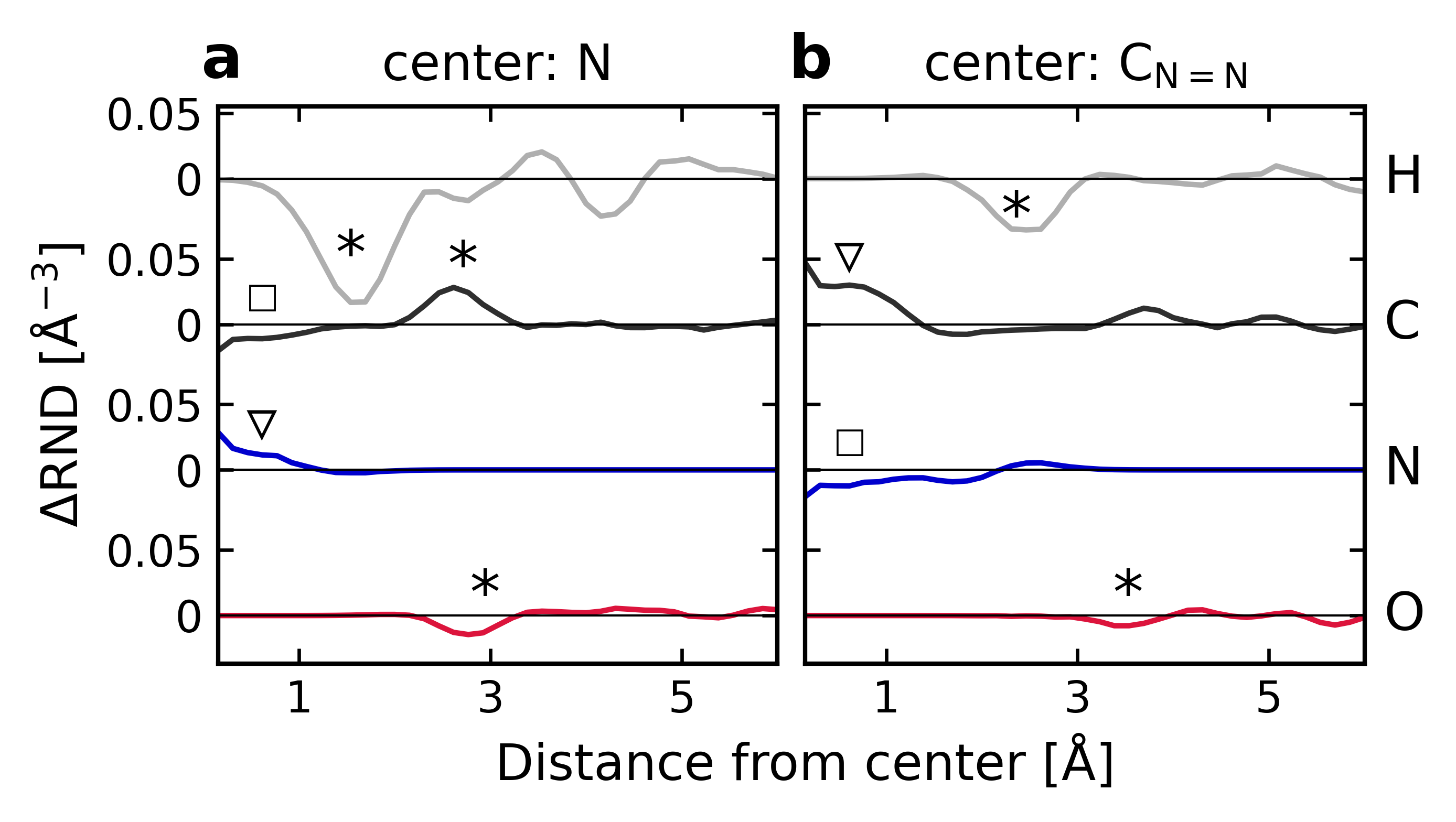}
    \caption{\label{fig:F3PBE} The corresponding presentation of Figure~3 of the main text, obtained using the PBE functional. }
\end{figure}

\begin{figure}[h!]
    \centering
    \includegraphics[width=0.5\columnwidth]{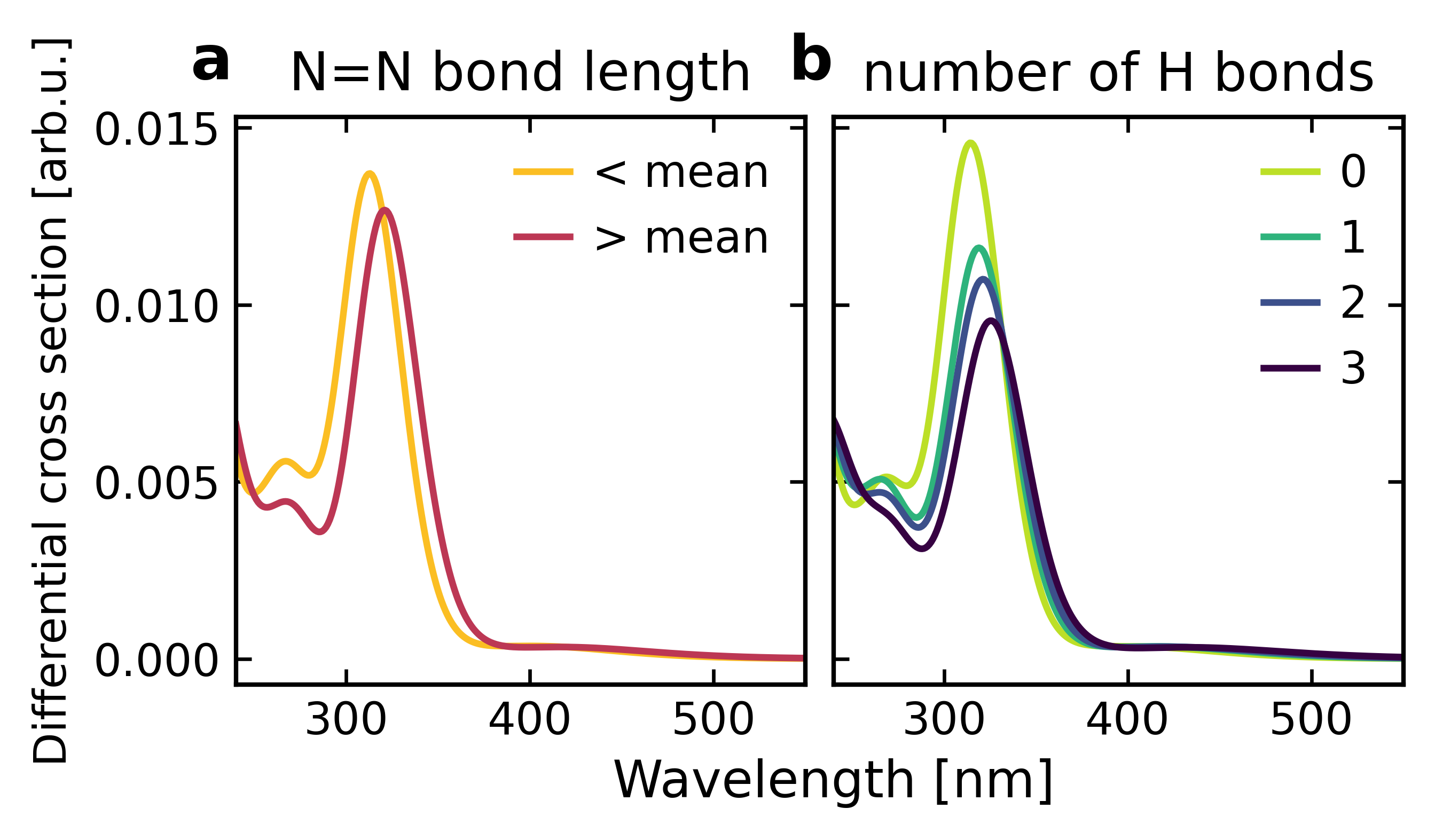}
    \caption{\label{fig:F4PBE} The corresponding presentation of Figure~4 of the main text, obtained using the PBE functional. }
\end{figure}

\end{document}